\documentclass[aps,preprint,nofootinbib,unsortedaddress]{revtex4}%
\usepackage{amssymb}
\usepackage{amsfonts}
\usepackage{amsmath}
\usepackage{graphicx}
\usepackage{hyperref}
\usepackage[usenames]{color}%
\providecommand{\U}[1]{\protect\rule{.1in}{.1in}}
\begin{document}
\preprint{ }
\title{Collective excitations of a charged Fermi superfluid in the BCS-BEC crossover}
\author{S. N. Klimin}
\affiliation{{TQC, Universiteit Antwerpen, Universiteitsplein 1, B-2610 Antwerpen, Belgium}}
\thanks{Author to whom any correspondence should be addressed. E-mail:
\href{mailto:sergei.klimin@uantwerpen.be}{sergei.klimin@uantwerpen.be}}
\author{J. Tempere}
\affiliation{{TQC, Universiteit Antwerpen, Universiteitsplein 1, B-2610 Antwerpen, Belgium}}
\altaffiliation{Also at: {Lyman Laboratory of Physics, Harvard University, USA}}
\author{T. Repplinger}
\affiliation{Laboratoire de Physique Th\'{e}orique, Universit\'{e} de Toulouse, CNRS, UPS,
31400, Toulouse, France}
\author{H. Kurkjian}
\affiliation{Laboratoire de Physique Th\'{e}orique, Universit\'{e} de Toulouse, CNRS, UPS,
31400, Toulouse, France}

\begin{abstract}
We consider collective excitations in the superfluid state of Fermi condensed
charged gases. The dispersion and damping of collective excitations at nonzero
temperatures are examined, and the coexistence and interaction of different
branches of collective excitations: plasma oscillations, pair-breaking Higgs
modes, and Carlson-Goldman phonon-like excitations are taken into account. The
path integral methods for superfluid Fermi gases and for Coulomb gas are
combined into a unified formalism that extends the Gaussian fluctuation
approximation to account for plasmonic modes. This approximation of Gaussian
pair and density fluctuations is able to describe all branches of collective
excitations existing in a charged superfluid. The spectra of collective
excitations are determined in two ways: from the spectral functions and from
the complex poles of the fluctuation propagator. A resonant avoided crossing
of different modes is shown. It is accompanied by resonant enhancement of the
response provided by the pair-breaking modes due to their interaction with
plasma oscillations. This may facilitate the experimental observation of the
pair-breaking modes.

\end{abstract}
\date{\today}
\maketitle

\section{Introduction \label{Intro}}

For decades, collective excitations in neutral and charged superfluids have
been the subject of great interest in condensed matter physics. Their
manifestations are found in a wide range of phenomena, from superconductors
and quantum gases to nuclear systems and neutron stars
\cite{Strinati,Kagan,Pekker,Shimano,Sedrakian}. Collective excitations are
important for both experiment and theory because they determine the response
spectra of condensed systems. Interest in collective excitations in superfluid
and superconducting systems has been reinforced by experiments to study their
response properties.

The present work focuses on collective excitations in charged superfluid Fermi
fluids and superconductors. Several branches of collective excitations are a
subject of the present study. The gapless soundlike mode
\cite{Anderson1958,Ohashi2003,Diener2008,PB-PRA}, called Anderson-Bogoliubov
mode, is well specified in neutral superfluid systems such as cold atomic
gases. In superconductors, this gapless mode is affected by the Coulomb
interaction and pushed up to the plasma mode. In the long-wavelength limit,
the plasma mode is gapped, and the size of the gap is the same both in the
superfluid/superconducting and in the normal state \cite{Anderson1958,Takada2}%
. Near the transition temperature $T_{c}$, the other gapless mode can appear
in BCS superconductors, discovered by Carlson and Goldman
\cite{Carlson,Takada1}.

The plasma mode is associated with oscillations of the particle density and is
therefore well resolved in the density response. For the pair field response,
the belonging of different branches of collective excitations to the pure
amplitude and phase responses is only asymptotically exact in the far BCS
limit. Plasma and Anderson-Bogoliubov collective excitations are revealed in
the pair field response through oscillations of the superfluid phase. There
exists also an amplitude mode due to oscillations of the pair field modulus,
attributed in many papers to the Higgs mechanism \cite{Pekker,Shimano} and
revisited recently in Ref. \cite{PB-PRL} as the \emph{pair-breaking} mode. In
Ref. \cite{Hoinka}, spectra of both Anderson-Bogoliubov and pair-breaking
modes have been experimentally studied for neutral atomic Fermi superfluids.
The pair-breaking collective excitations in superconductors were theoretically
predicted long ago \cite{Littlewood}, but have only recently been discovered
experimentally \cite{MatsunagaPRL}.

In this paper special attention is paid to the interaction between different
branches of collective excitations in charged Fermi superfluids. In the theory
of collective excitations in superconductors, plasma frequency is usually
assumed to be very large with respect to the superconducting gap. Here, we
focus on the other case, when they are comparable to each other. It may be
realized in strong-coupling superconductors where the BCS-BEC crossover regime
can exist, particularly in iron-based superconductors. Consequently, the treatment is
performed using methods suitable for the crossover. Also, collective
excitations in high-$T_{c}$ superconductors \cite{Deutscher} can reveal an
interaction of plasma and pair-field branches. The energy spectrum of charge
carriers in high-$T_{c}$ superconductors is substantially different from the
single-band 3D picture exploited in the present study, but the extension of
the formalism to a two-dimensional and multiband system is straightforward. It
is a subject of the future investigation.

In the BCS-BEC crossover, neither the plasma, nor the Anderson-Bogoliubov, nor
the pair-breaking mode can be attributed exactly to modulus or phase
responses, because of amplitude-phase mixing, which is also a subject of
attention in this treatment. Another important point of the study is to
investigate collective excitations at nonzero temperatures, where the spectrum
of excitations and the picture of an interplay of different branches is richer
than at zero temperature.

In the present work we apply the Gaussian fluctuation approach, which is well
established for a description of collective excitations in neutral superfluids
of cold atomic gases
\cite{Engelbrecht,Ohashi2003,Diener2008,PB-PRA,AllModes-PRA}. It can be
straightforwardly extended to charged superfluid Fermi gases by addition of a
Hubbard-Stratonovich field which describes oscillations of the particle
density \cite{Sharapov,Sharapov2} and is promising also for application to
superconductors. The Gaussian fluctuation method is equivalent to the
extraction of the excitation spectra from the linear response within the
random phase approximation (RPA) which is also frequently used in the theory
of collective excitations \cite{Anderson1958,Takada2,Takada1,PB-PRL}.

\section{Path integral approach \label{Theory}}

\subsection{Effective action and Gaussian fluctuation approximation
\label{EffAction}}

We consider a charged Fermi gas using the path-integral formalism. The
thermodynamic properties of an interacting Fermi gas are determined by the
partition function represented through the path integral over Grassmann field
variables $\left\{  \bar{\psi}_{\sigma},\psi_{\sigma}\right\}  $%
\begin{equation}
\mathcal{Z}=\int e^{-S}D\left[  \bar{\psi},\psi\right]  \label{PF}%
\end{equation}
with the action functional $S$,%
\begin{equation}
S=\int_{0}^{\beta}d\tau\int d\mathbf{r}\left[  \sum_{\sigma=\uparrow
,\downarrow}\bar{\psi}_{\sigma}\left(  \frac{\partial}{\partial\tau}%
+H-\mu\right)  \psi_{\sigma}+g\bar{\psi}_{\uparrow}\bar{\psi}_{\downarrow}%
\psi_{\downarrow}\psi_{\uparrow}\right]  +S_{C} \label{S}%
\end{equation}
where the model attractive interaction for the pairing channel is
expressed by a contact potential with the coupling constant $g<0$. Fermions
are assumed to have the spin $1/2$, $\sigma=\left(  \uparrow,\downarrow
\right)  $ are the spin projections. The part of the action describing the
Coulomb interaction is%
\begin{equation}
S_{C}=\frac{1}{2}\int_{0}^{\beta}d\tau\int d\mathbf{r}\int d\mathbf{r}%
^{\prime}U_{C}\left(  \mathbf{r}^{\prime}-\mathbf{r}\right)  \rho\left(
\mathbf{r}\right)  \rho\left(  \mathbf{r}^{\prime}\right)  \label{SC}%
\end{equation}
with the particle density%
\begin{equation}
\rho\left(  \mathbf{r}\right)  =\sum_{\sigma}\bar{\psi}_{\sigma}\left(
\mathbf{r}\right)  \psi_{\sigma}\left(  \mathbf{r}\right)
\end{equation}
and the Coulomb interaction potential%
\begin{equation}
U_{C}\left(  \mathbf{r}\right)  =\frac{e^{2}}{4\pi\epsilon_{0}\varepsilon
\left\vert \mathbf{r}\right\vert } \label{UC}%
\end{equation}
where $\epsilon_{0}$ is the permittivity of free space, $\varepsilon$ is the
high-frequency dielectric constant of a medium. In the absence of the Coulomb
interaction, (\ref{S}) turns into the widely used fermionic action
\cite{deMelo1993}.

We choose the units: $\hbar=1$, $m=1/2$ and $k_{F}\equiv\left(  3\pi
^{2}n\right)  ^{1/3}=1$, which leads to the equality $E_{F}=1$ for
free-fermion Fermi energy $E_{F}\equiv\frac{\hbar^{2}k_{F}^{2}}{2m}$. For a
charged gas, one more input parameter appears: the effective charge
$e/\sqrt{4\pi\epsilon_{0}\varepsilon}$. It can be expressed through the
dimensionless parameter, the Coulomb $\alpha_{0}$ having some analogy with the
Fr\"{o}hlich electron-phonon interaction constant $\alpha$:%
\begin{equation}
\alpha_{0}\equiv\frac{e^{2}}{4\pi\epsilon_{0}\varepsilon\hbar}\sqrt{\frac
{2m}{E_{F}}}.
\end{equation}
In these units, the bare\ plasma frequency $\omega_{p}=\sqrt{e^{2}%
n/\epsilon_{0}\varepsilon m}$ is expressed as $\omega_{p}=\sqrt{\left(
8/3\pi\right)  \alpha_{0}}$.

It is worth discussing how can the chosen model potential be relevant to
experiments on collective excitations in charged Fermi superfluids. In order
to clarify the subject of the study, we note that this term means
superconductors within the present work. Another class of systems which might
represent an interest for the application of the used theoretical method and
its future development, are ultracold ionic or atom-ionic gases
\cite{Tomza2019}. However, they do not yet realize a superfluid state for a
charged component, while the present treatment involves pair field as an
essential element. Therefore at the present state-of-art of experiment, other
charged fermion systems are not considered in the present investigation.

In rather early works \cite{deMelo1993,Ohashi2003}, the effective action
approach is used exploiting only the contact model pairing interaction,
assuming that observable effects of a complete true interaction potential can
be summarized through a single parameter, the effective scattering length. In
these works, the discussion was therefore performed in the context of both
superconductors and atomic superfluids. This approach however was not able to
describe the plasma branch of collective excitations, because it needs to
account for the Coulomb interaction explicitly.

For a Coulomb gas without pairing, the path-integral approach using the
effective action with the Coulomb interaction described by (\ref{SC}) and
(\ref{UC}) \cite{PopovBook} appears to be equivalent to the random phase
approximation. It effectively describes the plasma collective excitations in
the normal state of a gas of electrically charged fermions.

The straightforward way to consider collective excitations taking into account
both plasma oscillations and excitations of the pair field is to include the
Coulomb repulsion potential as a separate term in the model interaction
potential. The same model potential as in the present treatment, which is a
sum of pairing interaction and Coulomb potentials, has been already used in a
series of preceding publications, e. g., using the contact
\cite{Takada2,Wong1988} or finite-width \cite{Sharapov,Sharapov2} pairing
interaction potential, or an effective pairing potential provided by the
electron-phonon interaction \cite{Anderson1958}.

The repulsion between electrons is screened in superconductors, so that the
effective repulsion can differ from the Coulomb potential (\ref{UC}). It
should be noted however that the random phase approximation (which is
equivalent to the Gaussian fluctuation approximation applied below) leads, in
particular, to dynamic screening of the Coulomb interaction. As RPA leads to
the account of dynamic screening, it is logical to write down the Coulomb
potential in the non-screened form, to avoid double counting. See also the
discussion of Eq. (9) in Ref. \cite{Anderson1958}, where screening should be
taken into account for exchange terms. The exchange terms are neglected in the
present work as well as in the preceding paper \cite{Plasma-PRL}. They can be
non-negligible at strong-coupling. Here, we suggest that the calculated
spectra of collective excitations remain qualitatively correct in the BCS-BEC
crossover except maybe the BEC regime, which is beyond the scope of the
present work.

The other part of the interaction potential, which is responsible for the
pairing channel, is applied here in the form of a contact potential. As long
as calculated results are expressed in terms of the scattering length, a true
shape of the potential has no significance, because we have no aim to derive
this interaction from the first principles. The absence of such derivation of
course makes the theory more phenomenological than first-principle theories.
However its experimental relevance can be kept if we match the scattering
length with, for example, the ratio $\left.  \Delta\right\vert _{T=0}/E_{F}$,
which can be an independent \emph{input} parameter (what is done in figures
with numeric results of the present work). Also, the same method can be used
to compare different theoretical approaches to each other.

The effective bosonic action for a Fermi superfluid is obtained after
introducing two auxiliary bosonic fields: the pair field $\left(  \bar{\Psi
},\Psi\right)  $ and the density field $\Phi$, by adding them to the fermionic
action as follows:
\begin{equation}
S_{ext}=S+\int_{0}^{\beta}d\tau\int d\mathbf{r}\left[  -\frac{1}{g}\bar{\Psi
}\left(  \mathbf{r},\tau\right)  \Psi\left(  \mathbf{r},\tau\right)  +\frac
{1}{8\pi}\left(  \nabla\Phi\left(  \mathbf{r},\tau\right)  \right)
^{2}\right]  . \label{Sext}%
\end{equation}

The next step is the Hubbard-Stratonovich (HS) transformation, which shifts
the bosonic fields in order to remove the fermionic interaction in $S$. After
this, we use the Nambu representation of fermionic spinors, determined as%
\begin{equation}
\psi=\left(
\begin{array}
[c]{c}%
\psi_{1}\\
\psi_{2}%
\end{array}
\right)  =\left(
\begin{array}
[c]{c}%
\psi_{\uparrow}\\
\bar{\psi}_{\downarrow}%
\end{array}
\right)  . \label{Nambu}%
\end{equation}
The extended action (\ref{Sext}) after the HS shift is then%
\begin{equation}
S_{ext}^{\prime}=\int_{0}^{\beta}d\tau\int d\mathbf{r}\left[  \bar{\psi
}\left(  -\mathbb{G}^{-1}\right)  \psi-\frac{1}{g}\bar{\Psi}\Psi+\frac{1}%
{8\pi}\left(  \nabla\Phi\right)  ^{2}\right]  \label{SextB}%
\end{equation}
where the quadratic form with the inverse Nambu matrix is given by:%
\begin{equation}
-\mathbb{G}^{-1}=\left(
\begin{array}
[c]{cc}%
\frac{\partial}{\partial\tau}+H-\mu+i\sqrt{\alpha_{0}}\Phi & -\Psi\\
-\bar{\Psi} & \frac{\partial}{\partial\tau}-H+\mu-i\sqrt{\alpha_{0}}\Phi
\end{array}
\right)  . \label{NambuMatrix}%
\end{equation}
It differs from the analogous matrix for a neutral Fermi superfluid by the
presence of the terms provided by the Coulomb interaction, which are
proportional to the density field $\Phi$.

The integration over fermionic variables in the partition function with the
action $S_{ext}^{\prime}$ is performed formally exactly and leads to the
partition function expressed as the path integral over bosonic pair and
density fields,%
\begin{equation}
\mathcal{Z}\propto\int\mathcal{D}\bar{\Psi}\mathcal{D}\Psi\mathcal{D}\Phi
\exp\left(  -S_{\mathrm{eff}}\right)  \label{ZB}%
\end{equation}
with the effective bosonic action%
\begin{equation}
S_{\mathrm{eff}}=-\operatorname*{tr}\left[  \ln\left(  -\mathbb{G}%
^{-1}\right)  \right]  +\int_{0}^{\beta}d\tau\int d\mathbf{r}\left(  -\frac
{1}{g}\bar{\Psi}\Psi+\frac{1}{8\pi}\left(  \nabla\Phi\right)  ^{2}\right)  .
\label{Seff1}%
\end{equation}
For the subsequent derivation, we apply the Fourier representation for
fermionic and bosonic fields,%
\begin{subequations}
\begin{align}
\psi\left(  \mathbf{r},\tau\right)   &  =\frac{1}{\sqrt{V\beta}}%
\sum_{\mathbf{k}}\sum_{n=-\infty}^{\infty}\psi_{\mathbf{k},n}e^{i\mathbf{k}%
\cdot\mathbf{r}-i\omega_{n}\tau},\label{f1}\\
\Psi\left(  \mathbf{r},\tau\right)   &  =\frac{1}{\sqrt{V\beta}}%
\sum_{\mathbf{q}}\sum_{m=-\infty}^{\infty}\Psi_{\mathbf{q},m}e^{i\mathbf{q}%
\cdot\mathbf{r}-i\Omega_{m}\tau}\label{f2}\\
\Phi\left(  \mathbf{r},\tau\right)   &  =\frac{1}{\sqrt{V\beta}}%
\sum_{\mathbf{q}}\sum_{m=-\infty}^{\infty}\Phi_{\mathbf{q},m}e^{i\mathbf{q}%
\cdot\mathbf{r}-i\Omega_{m}\tau} \label{f3}%
\end{align}
with the fermion and boson Matsubara frequencies%
\end{subequations}
\begin{equation}
\omega_{n}=\frac{\left(  2n+1\right)  \pi}{\beta},\;\Omega_{n}=\frac{2n\pi
}{\beta}. \label{MF}%
\end{equation}
The effective bosonic action then takes the form%
\begin{equation}
S_{\mathrm{eff}}=-\operatorname*{tr}\left[  \ln\left(  -\mathbb{G}%
^{-1}\right)  \right]  -\sum_{\mathbf{q}}\sum_{m}\frac{1}{g}\bar{\Psi
}_{\mathbf{q},m}\Psi_{\mathbf{q},m}+\sum_{\mathbf{q}\neq0}\sum_{m}\frac{q^{2}%
}{8\pi}\Phi_{-\mathbf{q},-m}\Phi_{\mathbf{q},m}. \label{Seff}%
\end{equation}

In order to consider thermodynamics and response of the superfluid fermionic
system with the effective bosonic action (\ref{Seff}), the lowest-order
approximation is the saddle-point one, which determines macroscopic values of
the field variables $\left(  \Psi,\Phi\right)  $ from the least action
principle%
\begin{equation}
\frac{\delta S_{\mathrm{eff}}}{\delta\bar{\Psi}}=0,\qquad\frac{\delta
S_{\mathrm{eff}}}{\delta\Phi}=0. \label{LAP}%
\end{equation}
We apply trial saddle-point values of the pair and density fields to be
uniform in space, to consider collective excitations on top of a uniform
background. Also coordinate-dependent saddle-point solutions are possible
\cite{Anderson1958,PB-PRA}, but they are beyond the scope of the present work.
The uniform saddle-point value for the density field appears to be equal to
zero. Therefore we arrive at the gap equation for the saddle-point value of
the pair field, which takes the same form as for the neutral superfluid,%
\begin{equation}
\int\frac{d\mathbf{k}}{\left(  2\pi\right)  ^{3}}\left(  \frac{X\left(
E_{\mathbf{k}}\right)  }{2E_{\mathbf{k}}}-\frac{m}{k^{2}}\right)  +\frac
{m}{4\pi a_{s}}=0, \label{gap2}%
\end{equation}
where $E_{\mathbf{k}}=\sqrt{\xi_{\mathbf{k}}^{2}+\Delta^{2}}$ is the BCS
excitation energy and $\xi_{\mathbf{k}}=k^{2}-\mu$ the free-fermion energy.
{We note that in this formulation, our saddle-point approximation misses the
exchange scattering contributions which typically add a term of the form
$V(\mathbf{k}-\mathbf{k}^{\prime}){X\left(  E_{\mathbf{k}^{\prime}}\right)
}{2E_{\mathbf{k}^{\prime}}}$ to Eq.~\eqref{gap2} where $V(q)=q^{2}/8\pi$ is
the Fourier transform of the Coulomb potential.} The coupling constant of the
contact interaction is renormalized through the scattering length $a_{s}$
\cite{deMelo1993}:%
\begin{equation}
\frac{1}{g}=\frac{m}{4\pi a_{s}}-\int\frac{d^{3}k}{\left(  2\pi\right)  ^{3}%
}\frac{m}{k^{2}}, \label{g}%
\end{equation}
and $X\left(  E\right)  $ is the function%
\begin{equation}
X\left(  E\right)  =\tanh\left(  \frac{\beta E}{2}\right)  . \label{X}%
\end{equation}
The next approximation takes into account the fluctuations about the saddle
point,%
\begin{equation}
\Psi_{\mathbf{q},m}=\sqrt{V\beta}\Delta\delta_{\mathbf{q},0}\delta
_{m,0}+\varphi_{\mathbf{q},m} \label{fluct}%
\end{equation}
where $\varphi$ is the pair fluctuation field. Introducing the amplitude-phase
representation
\[
\varphi_{\mathbf{q},m}=\frac{\lambda_{\mathbf{q},m}+i\theta_{\mathbf{q},m}%
}{\sqrt{2}}\qquad\bar{\varphi}_{\mathbf{q},m}=\frac{\lambda_{\mathbf{q}%
,m}-i\theta_{\mathbf{q},m}}{\sqrt{2}}%
\]
with, respectively, amplitude and phase fluctuations $\lambda_{\mathbf{q},m}$
and $\theta_{\mathbf{q},m}$. These Fourier components correspond to real
amplitude and phase fields in time and space, so that $\bar{\lambda
}_{\mathbf{q},m}=\lambda_{-\mathbf{q},-m}$ and $\bar{\theta}_{\mathbf{q}%
,m}=\theta_{-\mathbf{q},-m}$. The resulting Gaussian pair and density
fluctuation (GPDF) action is given by:%
\begin{align}
S_{GPDF}  &  =\frac{1}{2}\sum_{\mathbf{q},m}\left(
\begin{array}
[c]{ccc}%
\lambda_{-\mathbf{q},-m} & \theta_{-\mathbf{q},-m} & \Phi_{-\mathbf{q},-m}%
\end{array}
\right) \nonumber\\
&  \times\left(
\begin{array}
[c]{ccc}%
K_{1,1} & K_{1,2} & K_{1,3}\\
-K_{1,2} & K_{2,2} & K_{2,3}\\
K_{1,3} & -K_{2,3} & K_{3,3}%
\end{array}
\right)  \left(
\begin{array}
[c]{c}%
\lambda_{\mathbf{q},m}\\
\theta_{\mathbf{q},m}\\
\Phi_{\mathbf{q},m}%
\end{array}
\right)  . \label{SGPDF4}%
\end{align}
The matrix elements of the pair-field (GPF) part of the GPDF action are
equivalent to those obtained in preceding works, e. g. Ref. \cite{Engelbrecht}%
, and read:
\begin{align}
K_{1,1}\left(  \mathbf{q},i\Omega_{m}\right)   &  =-\frac{1}{8\pi a_{s}}%
+\int\frac{d\mathbf{k}}{\left(  2\pi\right)  ^{3}}\left\{  \frac{1}{2k^{2}%
}+\frac{X\left(  E_{\mathbf{k}}\right)  }{4E_{\mathbf{k}}E_{\mathbf{k}%
+\mathbf{q}}}\right. \nonumber\\
&  \times\left[  \left(  \xi_{\mathbf{k}}\xi_{\mathbf{k}+\mathbf{q}%
}+E_{\mathbf{k}}E_{\mathbf{k}+\mathbf{q}}-\Delta^{2}\right)  \left(  \frac
{1}{i\Omega_{m}-E_{\mathbf{k}}-E_{\mathbf{k}+\mathbf{q}}}-\frac{1}{i\Omega
_{m}+E_{\mathbf{k}}+E_{\mathbf{k}+\mathbf{q}}}\right)  \right. \nonumber\\
&  \left.  \left.  +\left(  \xi_{\mathbf{k}}\xi_{\mathbf{k}+\mathbf{q}%
}-E_{\mathbf{k}}E_{\mathbf{k}+\mathbf{q}}-\Delta^{2}\right)  \left(  \frac
{1}{i\Omega_{m}-E_{\mathbf{k}+\mathbf{q}}+E_{\mathbf{k}}}-\frac{1}{i\Omega
_{m}-E_{\mathbf{k}}+E_{\mathbf{k}+\mathbf{q}}}\right)  \right]  \right\}  ,
\label{K11}%
\end{align}

\begin{align}
K_{2,2}\left(  \mathbf{q},i\Omega_{m}\right)   &  =-\frac{1}{8\pi a_{s}}%
+\int\frac{d\mathbf{k}}{\left(  2\pi\right)  ^{3}}\left\{  \frac{1}{2k^{2}%
}+\frac{X\left(  E_{\mathbf{k}}\right)  }{4E_{\mathbf{k}}E_{\mathbf{k}%
+\mathbf{q}}}\right. \nonumber\\
&  \times\left[  \left(  \xi_{\mathbf{k}}\xi_{\mathbf{k}+\mathbf{q}%
}+E_{\mathbf{k}}E_{\mathbf{k}+\mathbf{q}}+\Delta^{2}\right)  \left(  \frac
{1}{i\Omega_{m}-E_{\mathbf{k}}-E_{\mathbf{k}+\mathbf{q}}}-\frac{1}{i\Omega
_{m}+E_{\mathbf{k}}+E_{\mathbf{k}+\mathbf{q}}}\right)  \right. \nonumber\\
&  \left.  \left.  +\left(  \xi_{\mathbf{k}}\xi_{\mathbf{k}+\mathbf{q}%
}-E_{\mathbf{k}}E_{\mathbf{k}+\mathbf{q}}+\Delta^{2}\right)  \left(  \frac
{1}{i\Omega_{m}-E_{\mathbf{k}+\mathbf{q}}+E_{\mathbf{k}}}-\frac{1}{i\Omega
_{m}-E_{\mathbf{k}}+E_{\mathbf{k}+\mathbf{q}}}\right)  \right]  \right\}  ,
\label{K22}%
\end{align}

\begin{align}
K_{1,2}\left(  \mathbf{q},i\Omega_{m}\right)   &  =i\int\frac{d\mathbf{k}%
}{\left(  2\pi\right)  ^{3}}\frac{X\left(  E_{\mathbf{k}}\right)
}{4E_{\mathbf{k}}E_{\mathbf{k}+\mathbf{q}}}\nonumber\\
&  \times\left[  \left(  \xi_{\mathbf{k}}E_{\mathbf{k}+\mathbf{q}%
}+E_{\mathbf{k}}\xi_{\mathbf{k}+\mathbf{q}}\right)  \left(  \frac{1}%
{i\Omega_{m}-E_{\mathbf{k}}-E_{\mathbf{k}+\mathbf{q}}}+\frac{1}{i\Omega
_{m}+E_{\mathbf{k}}+E_{\mathbf{k}+\mathbf{q}}}\right)  \right. \nonumber\\
&  \left.  +\left(  \xi_{\mathbf{k}}E_{\mathbf{k}+\mathbf{q}}-E_{\mathbf{k}%
}\xi_{\mathbf{k}+\mathbf{q}}\right)  \left(  \frac{1}{i\Omega_{m}%
-E_{\mathbf{k}+\mathbf{q}}+E_{\mathbf{k}}}+\frac{1}{i\Omega_{m}-E_{\mathbf{k}%
}+E_{\mathbf{k}+\mathbf{q}}}\right)  \right]  , \label{K12}%
\end{align}%
\begin{equation}
K_{2,1}\left(  \mathbf{q},i\Omega_{m}\right)  =-K_{1,2}\left(  \mathbf{q}%
,i\Omega_{m}\right)  . \label{K21}%
\end{equation}

The other matrix elements have been derived in Refs. \cite{Plasma-PRL,Castin}.
Explicitly, they are:%
\begin{align}
K_{1,3}\left(  \mathbf{q},i\Omega_{m}\right)   &  =-i\sqrt{2\alpha_{0}}%
\Delta\int\frac{d\mathbf{k}}{\left(  2\pi\right)  ^{3}}\frac{X\left(
E_{\mathbf{k}}\right)  }{4E_{\mathbf{k}}E_{\mathbf{k}+\mathbf{q}}}\left(
\xi_{\mathbf{k}}+\xi_{\mathbf{k}+\mathbf{q}}\right) \nonumber\\
&  \times\left(  \frac{1}{i\Omega_{m}-E_{\mathbf{k}}-E_{\mathbf{k}+\mathbf{q}%
}}-\frac{1}{i\Omega_{m}+E_{\mathbf{k}}+E_{\mathbf{k}+\mathbf{q}}}\right)
\nonumber\\
&  \left.  +\frac{1}{i\Omega_{m}+E_{\mathbf{k}}-E_{\mathbf{k}+\mathbf{q}}%
}-\frac{1}{i\Omega_{m}-E_{\mathbf{k}}+E_{\mathbf{k}+\mathbf{q}}}\right)  ,
\end{align}

\begin{align}
K_{2,3}\left(  \mathbf{q},i\Omega_{m}\right)   &  =-\sqrt{2\alpha_{0}}%
\Delta\int\frac{d\mathbf{k}}{\left(  2\pi\right)  ^{3}}\frac{X\left(
E_{\mathbf{k}}\right)  }{4E_{\mathbf{k}}E_{\mathbf{k}+\mathbf{q}}}\nonumber\\
&  \times\left[  \left(  E_{\mathbf{k}+\mathbf{q}}+E_{\mathbf{k}}\right)
\left(  \frac{1}{i\Omega_{m}-E_{\mathbf{k}}-E_{\mathbf{k}+\mathbf{q}}}%
+\frac{1}{i\Omega_{m}+E_{\mathbf{k}}+E_{\mathbf{k}+\mathbf{q}}}\right)
\right. \nonumber\\
&  \left.  +\left(  E_{\mathbf{k}+\mathbf{q}}-E_{\mathbf{k}}\right)  \left(
\frac{1}{i\Omega_{m}+E_{\mathbf{k}}-E_{\mathbf{k}+\mathbf{q}}}+\frac
{1}{i\Omega_{m}-E_{\mathbf{k}}+E_{\mathbf{k}+\mathbf{q}}}\right)  \right]  ,
\end{align}
which satisfy the symmetry properties%
\begin{align}
K_{3,1}\left(  \mathbf{q},i\Omega_{m}\right)   &  =K_{1,3}\left(
\mathbf{q},i\Omega_{m}\right)  ,\\
K_{3,2}\left(  \mathbf{q},i\Omega_{m}\right)   &  =-K_{2,3}\left(
\mathbf{q},i\Omega_{m}\right)  ,
\end{align}
and the diagonal matrix element is%
\begin{align}
K_{3,3}\left(  \mathbf{q},i\Omega_{m}\right)   &  =\frac{q^{2}}{4\pi}%
-\alpha_{0}\int\frac{d\mathbf{k}}{\left(  2\pi\right)  ^{3}}\frac{X\left(
E_{\mathbf{k}}\right)  }{2E_{\mathbf{k}}E_{\mathbf{k}+\mathbf{q}}}\nonumber\\
&  \times\left(  \frac{E_{\mathbf{k}}E_{\mathbf{k}+\mathbf{q}}-\xi
_{\mathbf{k}}\xi_{\mathbf{k}+\mathbf{q}}+\Delta^{2}}{i\Omega_{m}%
-E_{\mathbf{k}}-E_{\mathbf{k}+\mathbf{q}}}+\frac{E_{\mathbf{k}}E_{\mathbf{k}%
+\mathbf{q}}+\xi_{\mathbf{k}}\xi_{\mathbf{k}+\mathbf{q}}-\Delta^{2}}%
{i\Omega_{m}-E_{\mathbf{k}}+E_{\mathbf{k}+\mathbf{q}}}\right. \nonumber\\
&  \left.  -\frac{E_{\mathbf{k}}E_{\mathbf{k}+\mathbf{q}}+\xi_{\mathbf{k}}%
\xi_{\mathbf{k}+\mathbf{q}}-\Delta^{2}}{i\Omega_{m}+E_{\mathbf{k}%
}-E_{\mathbf{k}+\mathbf{q}}}-\frac{E_{\mathbf{k}}E_{\mathbf{k}+\mathbf{q}}%
-\xi_{\mathbf{k}}\xi_{\mathbf{k}+\mathbf{q}}+\Delta^{2}}{i\Omega
_{m}+E_{\mathbf{k}}+E_{\mathbf{k}+\mathbf{q}}}\right)  . \label{K33}%
\end{align}

\subsection{Analytic determination of collective excitation spectra
\label{AnCont}}

Energies and damping factors of collective excitations can be determined
through complex poles of the GPDF propagator, following to the procedure
proposed by Nozi\`{e}res \cite{Nozieres}. The formalism remains the same as in
Refs. \cite{PB-PRL,PB-PRA,AllModes-PRA}. For completeness, we reproduce here
its main steps. Formally, the complex poles of the GPDF propagator $\left(
\det\mathbb{K}\right)  ^{-1}$ are determined by the equation in the complex
$z$ plane,
\begin{equation}
\det\mathbb{K}\left(  z\right)  =0. \label{detKeq0}%
\end{equation}
Due to the branch cut at the real axis, these poles can become visible only
after the analytic continuation of the propagator through the branch cut to
the next sheet of the Riemann surface. Otherwise, they are hidden by the
branch cut as behind a mirror wall.

The analytic continuation is performed using the spectral density function. At
the real axis, it is determined by:
\begin{equation}
\rho_{K}\left(  \omega\right)  =\lim_{\delta\rightarrow0}\frac{\det
\mathbb{K}\left(  \omega+i\delta\right)  -\det\mathbb{K}\left(  \omega
-i\delta\right)  }{2\pi i}. \label{rho}%
\end{equation}
The spectral density is analytic on the real axis except maybe a finite number
of points. In any chosen interval between these points it can be
straightforwardly continued analytically to complex $z$ plane. Let us label
these intervals by the index $n$, and denote the spectral density in each
interval as $\rho_{K}^{\left(  n\right)  }\left(  \omega\right)  $, so that
the analytic continuation of the spectral density from each interval is
$\rho_{K}^{\left(  n\right)  }\left(  z\right)  $. The analytic continuation
$\det\mathbb{K}_{\downarrow}\left(  z\right)  $ of $\det\mathbb{K}\left(
z\right)  $ through the branch cut in the $n$-th interval is then%
\begin{equation}
\det\mathbb{K}_{\downarrow}^{\left(  n\right)  }\left(  z\right)  =\left\{
\begin{array}
[c]{cc}%
\det\mathbb{K}\left(  z\right)  , & \operatorname{Im}z>0,\\
\det\mathbb{K}\left(  z\right)  +2\pi i\rho_{K}^{(n)}\left(  z\right)  , &
\operatorname{Im}z<0.
\end{array}
\right.  \label{ancont}%
\end{equation}
The set of eigenfrequencies and damping factors for collective modes is
therefore determined by roots of the equations
\begin{equation}
\det\mathbb{K}_{\downarrow}^{\left(  n\right)  }\left(  z\right)  =0,
\label{detKeq0a}%
\end{equation}
for all intervals $n$. The bounds between different intervals in which
$\rho_{K}\left(  \omega\right)  $ is analytic are the angular points. They
indicate a change in the configuration of the resonant wave vectors for one of
the resonance conditions,
\begin{equation}
E_{\mathbf{k}-\frac{\mathbf{q}}{2}}+E_{\mathbf{k}+\frac{\mathbf{q}}{2}}%
=\omega,\quad\left\vert E_{\mathbf{k}-\frac{\mathbf{q}}{2}}-E_{\mathbf{k}%
+\frac{\mathbf{q}}{2}}\right\vert =\omega\label{eq1}%
\end{equation}
The case $\omega>0$ is considered, without loss of generality.%

\begin{figure}[ptbh]%
\centering
\includegraphics[
height=2.7544in,
width=3.5129in
]%
{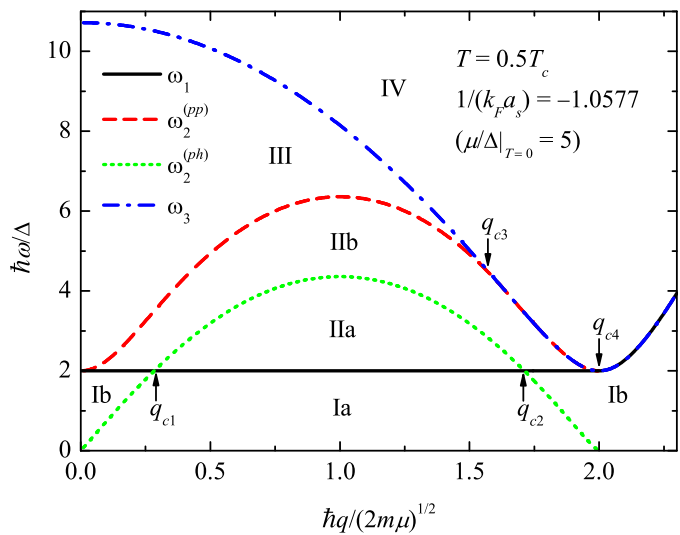}%
\caption{Angular-point frequencies for the analytic continuation of the GPF
matrix elements for $\left.  \left(  \mu/\Delta\right)  \right\vert _{T=0}=5$
and $T=0.5T_{c}$ (after Ref. \cite{AllModes-PRA}). The areas between curves
determine intervals for the analytic continuation as described in the text.
The arrows show values of momentum $q_{c1},\ldots q_{c4}$ at which different
angular-point frequencies coincide.}%
\label{fig:AngularPoints}%
\end{figure}

Fig. \ref{fig:AngularPoints} shows an example of angular points and,
correspondingly, intervals for the analytic continuation. A detailed
description of them can be found in Ref. \cite{AllModes-PRA}. Briefly, the
angular-point frequency $\omega_{1}$ is the pair-breaking continuum edge. The
angular points $\omega_{2}^{\left(  pp\right)  }$ (particle-particle) and
$\omega_{2}^{\left(  ph\right)  }$ (particle-hole) correspond to,
respectively, local extrema of the energies $E_{\mathbf{k}+\frac{\mathbf{q}%
}{2}}+E_{\mathbf{k}-\frac{\mathbf{q}}{2}}$ and $E_{\mathbf{k}+\frac
{\mathbf{q}}{2}}-E_{\mathbf{k}-\frac{\mathbf{q}}{2}}$ as functions of $k$ at
$\mathbf{k}\parallel\mathbf{q}$. The frequency $\omega_{3}$ is the energy of
the BCS pair $E_{\mathbf{k}-\frac{\mathbf{q}}{2}}+E_{\mathbf{k}+\frac
{\mathbf{q}}{2}}$ at zero momentum $k$. The angular point $\omega_{2}^{\left(
ph\right)  }$ has significance for $T\neq0$ and $T\neq T_{c}$, where the terms
with the particle-hole energy denominators bring nonzero contributions to the
GPDF matrix elements.

Because of multiplicity of intervals, some roots coming from the analytic
continuation through different intervals may be physically equivalent, i. e.,
correspond physically to the same modes. In this case, a selection of
appropriate roots must be performed using physical reasoning, as discussed
below in Sec. \ref{Results}.

\subsection{Response functions \label{RespFuns}}

The spectra of collective excitations can be rather qualitatively but reliably
extracted from spectral functions for several types of the response. Also,
spectral functions give an information on relative magnitudes of peaks
provided by different branches of collective excitations. Here, we begin with
two spectral functions for the pair field response: the modulus-modulus and
phase-phase spectral function, determined using the bosonic Green's functions
in the Matsubara representation,%
\begin{align}
\mathcal{G}_{\lambda\lambda}\left(  \mathbf{q},i\Omega_{m}\right)   &
=-\left\langle \lambda_{-\mathbf{q}-m}\lambda_{\mathbf{q}m}\right\rangle
_{S_{GPDF}},\label{GLL}\\
\mathcal{G}_{\theta\theta}\left(  \mathbf{q},i\Omega_{m}\right)   &
=-\left\langle \theta_{-\mathbf{q}-m}\theta_{\mathbf{q}m}\right\rangle
_{S_{GPDF}}. \label{GYY}%
\end{align}
The spectral functions are obtained after the analytic continuation from a
sequence of bosonic Matsubara frequencies to the complex plane near the real
axis,%
\begin{align}
\chi_{\lambda\lambda}\left(  \mathbf{q},\omega\right)   &  =-\frac{1}{\pi
}\operatorname{Im}\mathcal{G}_{\lambda\lambda}\left(  \mathbf{q},\omega
+i0^{+}\right)  ,\label{XiLL}\\
\chi_{\theta\theta}\left(  \mathbf{q},\omega\right)   &  =-\frac{1}{\pi
}\operatorname{Im}\mathcal{G}_{\theta\theta}\left(  \mathbf{q},\omega
+i0^{+}\right)  . \label{XiYY}%
\end{align}
Because the pair-density fluctuation action is quadratic, the bosonic averages
are explicitly calculated in a standard way. The resulting spectral weight
functions are then given by:%
\begin{align}
\chi_{\lambda\lambda}\left(  \mathbf{q},\omega\right)   &  =\frac{1}{\pi
}\operatorname{Im}\frac{K_{2,2}\left(  \mathbf{q},\omega+i0^{+}\right)
K_{3,3}\left(  \mathbf{q},\omega+i0^{+}\right)  +K_{2,3}^{2}\left(
\mathbf{q},\omega+i0^{+}\right)  }{\det\mathbb{K}\left(  \mathbf{q}%
,\omega+i0^{+}\right)  },\label{XLL}\\
\chi_{\theta\theta}\left(  \mathbf{q},\omega\right)   &  =\frac{1}{\pi
}\operatorname{Im}\frac{K_{1,1}\left(  \mathbf{q},\omega+i0^{+}\right)
K_{3,3}\left(  \mathbf{q},\omega+i0^{+}\right)  -K_{1,3}^{2}\left(
\mathbf{q},\omega+i0^{+}\right)  }{\det\mathbb{K}\left(  \mathbf{q}%
,\omega+i0^{+}\right)  }. \label{XYY}%
\end{align}

The density-density response function $\chi_{\rho\rho}$ is determined as%
\begin{equation}
\chi_{\rho\rho}\left(  \mathbf{q},\omega\right)  =-\frac{1}{\pi}%
\operatorname{Im}\mathcal{G}_{\rho}\left(  \mathbf{q},\omega+i0^{+}\right)
\label{hi}%
\end{equation}
through the Green's function $\mathcal{G}_{\rho}\left(  \mathbf{q},z\right)  $
which is the analytic continuation to the complex $z$ plane of the Matsubara
Green's function for Fourier components of the fermion density%
\begin{equation}
\mathcal{G}_{\rho}\left(  \mathbf{q},i\Omega_{m}\right)  \equiv-\left\langle
\rho_{-\mathbf{q},-m}\rho_{\mathbf{q},m}\right\rangle _{S}. \label{Grho1}%
\end{equation}
The averages in (\ref{Grho1}) are obtained using the generating functional
with the auxiliary source field $\upsilon$,%
\begin{equation}
\Xi_{\rho}\left[  \upsilon\right]  =\left\langle \exp\left[  \sum
_{\mathbf{q},m}\upsilon_{-\mathbf{q},-m}\rho_{\mathbf{q},m}\right]
\right\rangle _{S}, \label{Xi2}%
\end{equation}
using the relation
\begin{equation}
\mathcal{G}_{\rho}\left(  \mathbf{q},i\Omega_{m}\right)  \equiv-\left.
\frac{\partial^{2}\Xi_{\rho}\left[  \upsilon\right]  }{\partial\upsilon
_{-\mathbf{q},-m}\partial\upsilon_{\mathbf{q},m}}\right\vert _{\upsilon=0}.
\label{Diff}%
\end{equation}

The terms in (\ref{Xi2}) containing the source field, are included to the
fermionic action (\ref{S}), only resulting in addition of $\upsilon$ to the
chemical potential. Therefore the generating functional (\ref{Xi2}) is
calculated analytically in the same way as in Subsec. \ref{EffAction}:
performing the HS transformation and integrating over fermionic fields. As a
result, we arrive at the extended effective fluctuation action,%
\begin{align}
S_{GPDF}^{\prime}  &  =S_{GPF}+\frac{1}{2}\sum_{\mathbf{q},m}\left[
\frac{q^{2}}{4\pi}\Phi_{\mathbf{q},m}\Phi_{-\mathbf{q},-m}\right. \nonumber\\
&  +\left(  K_{3,3}-\frac{q^{2}}{4\pi}\right)  \left(  \Phi_{\mathbf{q}%
,m}+\frac{i}{\sqrt{\alpha_{0}}}\upsilon_{\mathbf{q},m}\right)  \left(
\Phi_{-\mathbf{q},-m}+\frac{i}{\sqrt{\alpha_{0}}}\upsilon_{-\mathbf{q}%
,-m}\right) \nonumber\\
&  \left.  +2\left(  K_{1,3}\lambda_{-\mathbf{q},-m}+K_{2,3}\theta
_{-\mathbf{q},-m}\right)  \left(  \Phi_{\mathbf{q},m}+\frac{i}{\sqrt
{\alpha_{0}}}\upsilon_{\mathbf{q},m}\right)  \right]  . \label{SGPDF1}%
\end{align}
Here, $S_{GPF}$ is the part of the GPDF action, which only describes
\emph{pair} fluctuations, and used in the theory of neutral Fermi superfluids,%
\begin{equation}
S_{GPF}=\frac{1}{2}\sum_{\mathbf{q},m}\left(
\begin{array}
[c]{cc}%
\lambda_{-\mathbf{q},-m} & \theta_{-\mathbf{q},-m}%
\end{array}
\right)  \left(
\begin{array}
[c]{cc}%
K_{1,1} & K_{1,2}\\
-K_{1,2} & K_{2,2}%
\end{array}
\right)  \left(
\begin{array}
[c]{c}%
\lambda_{\mathbf{q},m}\\
\theta_{\mathbf{q},m}%
\end{array}
\right)  . \label{SGPFa}%
\end{equation}
The relation (\ref{SGPDF1}) allows us to express the Green's function
(\ref{Grho1}) defined originally through the fermionic average (\ref{Grho1}),
in terms of averages of bosonic fields,%
\begin{align}
\mathcal{G}_{\rho}\left(  \mathbf{q}i\Omega_{m}\right)   &  =\frac{q^{2}%
/4\pi-K_{3,3}}{\alpha_{0}}+\frac{\left(  q^{2}/4\pi-K_{3,3}\right)  ^{2}%
}{\alpha_{0}}\left\langle \Phi_{\mathbf{q}m}\Phi_{-\mathbf{q}-m}\right\rangle
_{S_{GPDF}}\nonumber\\
&  +\frac{1}{\alpha_{0}}K_{1,3}^{2}\left\langle \lambda_{\mathbf{q}m}%
\lambda_{-\mathbf{q}-m}\right\rangle _{S_{GPDF}}-\frac{1}{\alpha_{0}}%
K_{2,3}^{2}\left\langle \theta_{\mathbf{q}m}\theta_{-\mathbf{q}-m}%
\right\rangle _{S_{GPDF}}\nonumber\\
&  +\frac{\left(  4\pi K_{3,3}-q^{2}\right)  }{2\pi\alpha_{0}}\left(
K_{1,3}\left\langle \lambda_{-\mathbf{q}-m}\Phi_{\mathbf{q}m}\right\rangle
_{S_{GPDF}}+K_{2,3}\left\langle \theta_{-\mathbf{q}-m}\Phi_{\mathbf{q}%
m}\right\rangle _{S_{GPDF}}\right) \nonumber\\
&  -\frac{2}{\alpha_{0}}K_{1,3}K_{2,3}\left\langle \theta_{\mathbf{q}m}%
\lambda_{-\mathbf{q}-m}\right\rangle _{S_{GPDF}}. \label{GrhoBF}%
\end{align}
As a result, after the calculation of bosonic averages, we arrive at a
remarkably compact expression%
\begin{equation}
\mathcal{G}_{\rho}\left(  \mathbf{q},i\Omega_{m}\right)  =\frac{q^{2}}%
{4\pi\alpha_{0}}\left(  \frac{q^{2}}{4\pi}\frac{\det\mathbb{K}_{GPF}\left(
\mathbf{q},i\Omega_{m}\right)  }{\det\mathbb{K}\left(  \mathbf{q},i\Omega
_{m}\right)  }-1\right)  , \label{CF}%
\end{equation}
where $\det\mathbb{K}_{GPF}$ is the determinant of the GPF $2\times2$ matrix,%
\[
\det\mathbb{K}_{GPF}=K_{1,1}K_{2,2}+K_{1,2}^{2},
\]
and $\det\mathbb{K}$ is the determinant of the whole fluctuation matrix
$\mathbb{K}$:%
\begin{equation}
\det\mathbb{K}=K_{3,3}\det\mathbb{K}_{GPF}+K_{1,1}K_{2,3}^{2}-K_{1,3}%
^{2}K_{2,2}+2K_{1,2}K_{1,3}K_{2,3}. \label{detP}%
\end{equation}
Consequently, according to (\ref{hi}), the density-density spectral weight
function is:%
\begin{equation}
\chi_{\rho\rho}\left(  \mathbf{q},\omega\right)  =-\frac{q^{4}}{16\pi
^{3}\alpha_{0}}\operatorname{Im}\left(  \frac{\det\mathbb{K}_{GPF}\left(
\mathbf{q},\omega+i0^{+}\right)  }{\det\mathbb{K}\left(  \mathbf{q}%
,\omega+i0^{+}\right)  }\right)  . \label{RF1}%
\end{equation}
In the limit $\alpha_{0}\rightarrow0$, this density-density spectral weight
function continuously turns to that for a neutral Fermi superfluid.

\section{Collective excitations at zero temperature \label{Results}}

We now present our results on collective excitations at $T=0$, in which case
frequencies below $2\Delta$ are free from damping. The high temperature case
($T\rightarrow T_{c}$), which exhibits a drastically different collective
physics, is treated in Sec.~\ref{sec:tc}. We use two complementary methods to
identify collective resonances: (i) the semianalytic determination of the
poles of the GPDF propagator Eq.~\eqref{detKeq0a}, which provides an
unambiguous determination of the eigenfrequency and damping rate of collective
modes, and (ii) a visual identification of {the resonances in the} spectral
functions Eqs.~(\ref{XLL},\ref{XYY},\ref{hi}), to gain a more phenomenological
view of collective effects in experimental observables. We stress that the
analytic continuation method allows for a more direct identification of the
prominent features of the spectrum, which is particularly useful when the
parameter space (in our case the 4D space spanned by $q/k_{F},1/k_{F}%
a,T/T_{c}$ and $\omega_{p}/\Delta$) is large. In the past \cite{PB-PRL,PB-PRA}%
, it was shown that the poles in the analytic continuation (together with
their residues) are usually a good summary of the behavior of the response
functions, even in the unconventional case where these pole have a large
imaginary part, comparable to their eigenfrequency. However, such highly
damped and distorted resonances are not elementary excitations strictly
speaking \cite{AGDBook}.%

\begin{figure}[tbh]%
\centering
\includegraphics[
height=4.1122in,
width=4.6561in
]%
{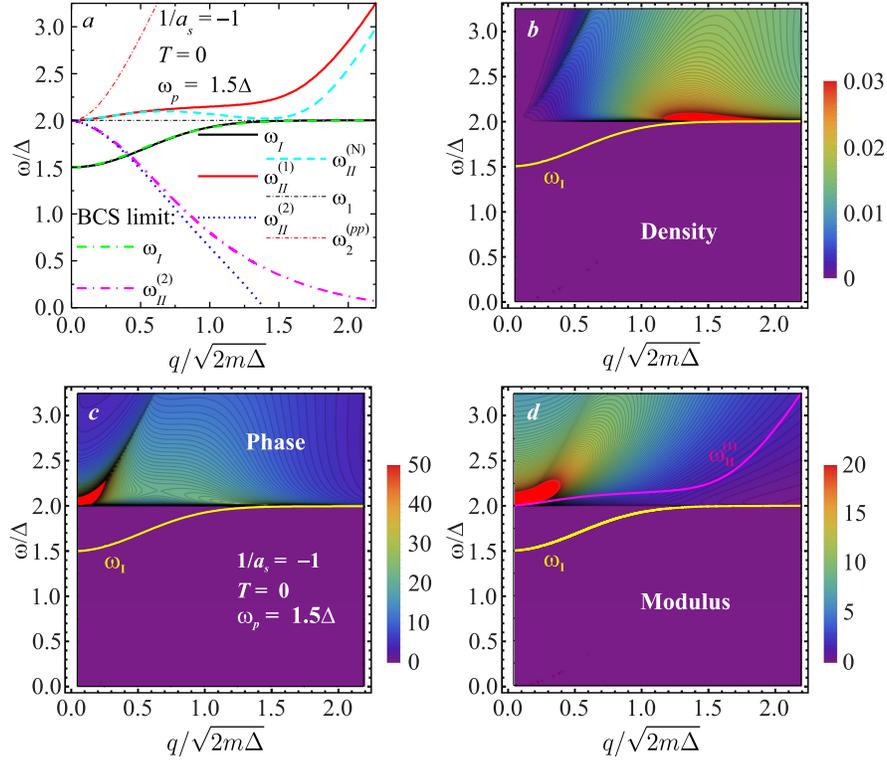}%
\caption{$(a)$ Eigenfrequencies given by real parts of complex roots of
$\text{det}\,\mathbb{K}_{\downarrow}$ (solid curves) at $1/k_{F}a_{s}=-1$
($\left.  \Delta\right\vert _{T=0}\approx0.2084E_{F}$), $\omega_{p}=1.5\Delta
$, $T=0$ and in function of the excitation wave vector $q$. The dispersions
are compared to their values in the BCS limit (dot-dashed curves) obtained
from Ref.~\cite{Plasma-PRL} with $q\xi=\sqrt{E_{F}/\Delta}\times
q/\sqrt{2m\Delta}$. The blue dashed curve shows the dispersion of the
pair-breaking mode for a neutral superfluid $\omega_{II}^{\left(  N\right)  }%
$. $(b,c,d)$ Contour plots of $\chi_{\rho\rho}$, $\chi_{\theta\theta}$ and
$\chi_{\lambda\lambda}$ in the same regime overlaid by the plasma frequency
$\omega_{I}$ from the complex root of $\mathbb{K}_{\downarrow}$ below the
transition temperature. The contour plot $\left(  d\right)  $ of the spectral
weight function for the modulus response is also overlaid by the pair-breaking
mode frequency $\omega_{II}^{\left(  1\right)  }$.}%
\label{fig:T0q}%
\end{figure}

In Fig. \ref{fig:T0q}, we compare the roots of $\det\mathbb{K}$ to the contour
plots of the spectral functions (restricting ourselves to the diagonal
modulus-modulus, phase-phase and density-density responses). We fix the plasma
frequency to $\omega_{p}=1.5\Delta$, the interaction strength to the BCS
regime ($1/k_{F}a=-1$ or equivalently $\Delta\approx0.2084E_{F}$), and we vary
the pair momentum $q$ to explore the dispersion of the modes. The dispersions
are overall similar to what was found in the BCS limit \cite{Plasma-PRL}
(reminded by the dashed-dotted curves in Fig.~\ref{fig:T0q}a).

\paragraph{Plasma branch}


Below the pair-breaking continuum, $\text{det}\mathbb{K}_{\downarrow}$ has an
undamped root $\omega_{\mathrm{I}}$ (black solid curve in Fig.~\ref{fig:T0q}a)
which produces a Dirac peak in the response functions (overlayed yellow curves
in Figs.~\ref{fig:T0q} b, c, d). This mode departs from $\omega_{p}$ at $q=0$,
such that at low $q$ it can be attributed without ambiguity to the plasma
mode. We remind that the existence of a plasma mode in $q=0$ and
$\omega=\omega_{p}$ is guaranteed by a sum rule \cite{Takada1}. In accordance
with the calculation by Anderson \cite{Anderson1958}, this demonstrates that
the gapless Goldstone mode disappears (at zero temperature at least) due to
long-range Coulomb interactions in a charged superfluid. For the chosen value
of $\omega_{p}$, the plasma mode has a positive dispersion but we remind that
when $\omega_{p}$ approaches $2\Delta$ from below, the dispersion becomes
nontrivial, particularly showing a downward dispersion with a minimum at some
nonzero $q$ \cite{Plasma-PRL}. At larger $q$, the plasma branch does not
enters smoothly into the pair-breaking continuum but splits into multiple
peaks both above and below $2\Delta$. In particular, it generates a sharp peak
at frequency $2\Delta^{+}$ (fairly constant in function of $q$), well visible
in Fig.~\ref{fig:T0q}\emph{b}, and partially explained by the presence of a
root $z_{\mathrm{II}}^{(2)}$ of $\det\mathbb{K}_{\downarrow}$ in window II
(blue dotted curve in Fig.~\ref{fig:T0q}\emph{a}). In the BCS limit, this root
belongs to the density-phase sector (it solves $K_{1,1}K_{3,3}-K_{1,3}^{2}%
=0$), contrarily to the pair-breaking mode $z_{\mathrm{II}}^{(1)}$ (red solid
curve in Fig.~\ref{fig:T0q}\emph{a}) which belongs to the modulus sector. We
note that a multiple resonance also appears in the phase-phase response
(Fig.~\ref{fig:T0q}\emph{c}), with a secondary peak visible around $\omega
_{2}^{(pp)}$ at low $q$. Absent in the neutral case, this peak is a signature
of long-range interactions. Remarkably, due to non-vanishing modulus-phase and
modulus-density couplings ($K_{1,2}$ and $K_{2,3}$ respectively), the plasma
mode also appears in the modulus-modulus channel, which was not the case in
the BCS limit. In the interaction regime considered here, it has a small
spectral weight at low $q$.

\paragraph{Pair-breaking branch}

A pair-breaking (\textquotedblleft Higgs\textquotedblright) mode
($z_{\mathrm{II}}^{(1)}$ or in short $z_{\mathrm{pb}}$) is found in the
analytic continuation through window II. At low-$q$ its dispersion is
qualitatively the same as in a neutral Fermi superfluid \cite{PB-PRL,PB-PRA},
departing quadratically from $2\Delta$. The branch (overlayed in purple in
Fig.~\ref{fig:T0q}\emph{d}) dominates the modulus-modulus response inside the
pair-breaking continuum. The most remarkable difference between
Fig.~\ref{fig:T0q} and the neutral case \cite{PB-PRL,Plasma-PRL} is the
notable rise in the frequency of the modulus mode for $q/\sqrt{2m\Delta}>1$
(compare the red and blue dashed curves), which we interpret as due to a
repulsive interaction with the plasma mode. Since this interaction is carried
by the matrix elements $K_{2,3}$ (modulus-density) and $K_{1,2}$
(modulus-phase), it vanishes in the BCS limit, such that the modulus mode in
the neutral and charged cases coincide in this limit.

\begin{figure}[tbh]%
\centering
\includegraphics[
height=5.5357in,
width=4.2938in
]%
{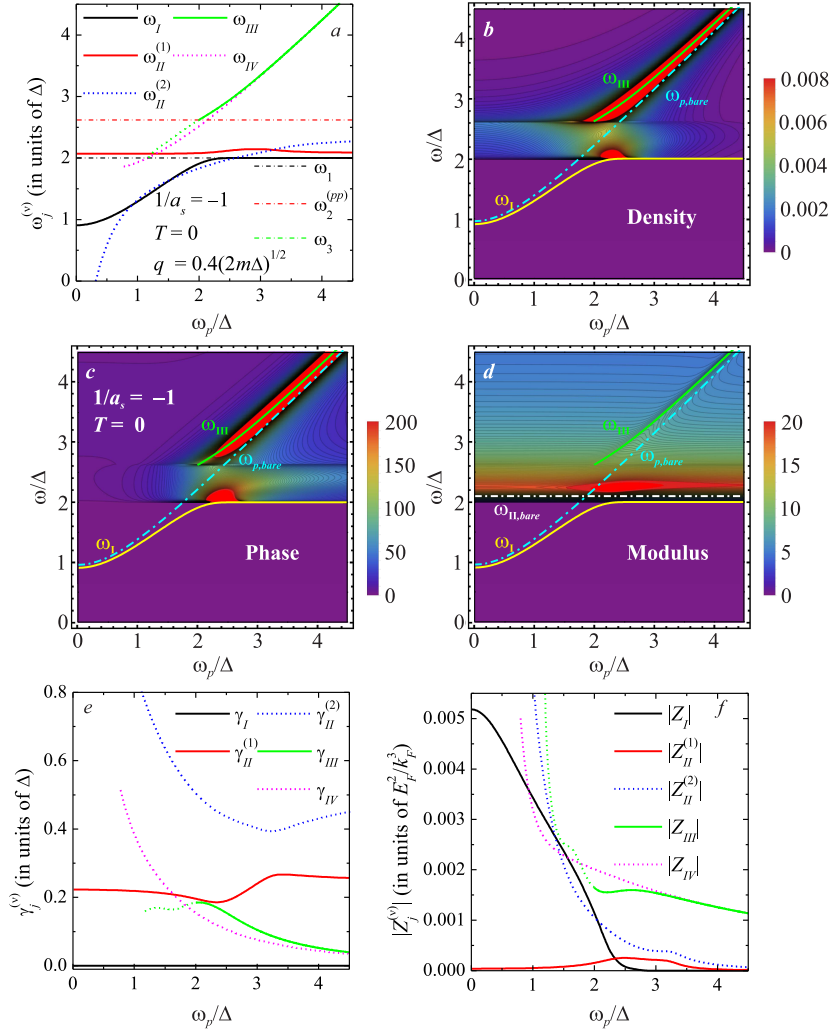}%
\caption{$(a)$ Eigenfrequencies provided by complex roots of $\text{det}%
\,\mathbb{K}_{\downarrow}$ (solid curves) at $1/k_{F}a_{s}=-1$ and $T=0$, this
time in function of $\omega_{p}$ at fixed $q/\sqrt{2m\Delta}=0.4$. $(b,c,d)$
Contour plots of $\chi_{\rho\rho}$, $\chi_{\theta\theta}$ and $\chi
_{\lambda\lambda}$ in the same regime, overlaid with the bare Higgs
eigenfrequency ($\omega_{\mathrm{II}}(\omega_{p}=0)=2.1\Delta$) and bare
plasma branch $\sqrt{\omega_{p}^{2}+c^{2}q^{2}}$ (with $cq/\Delta=0.95759$
here). $\left(  e,f\right)  $ Respectively, damping factors and moduli of
residues provided by complex roots of $\text{det}\,\mathbb{K}_{\downarrow}$.}%
\label{fig:T0wp}%
\end{figure}

To further illustrate the mixing of pair-breaking and plasma mode when their
frequencies are in resonance, we show in Fig.~\ref{fig:T0wp} the poles and the
spectral functions in function of the plasma frequency at fixed $q=0.4\sqrt
{2m\Delta}$. At this wavevector, the bare Higgs eigenfrequency (calculated at
$\omega_{p}=0$) is $\omega_{\mathrm{pb}}=2.1\Delta$. We overlay to the contour
plots the \textquotedblleft bare\textquotedblright\ frequency of the plasma
branch $\sqrt{\omega_{p}^{2}+c^{2}q^{2}}$ ($c$ is the sound velocity of the
neutral gas) which accurately predicts the location of the resonance away from
the interval $[2\Delta,\omega_{2}^{(pp)}]$. The interaction between plasma and
modulus modes reveals itself through the intensity increase of the resonance
in the $[2\Delta,\omega_{2}^{(pp)}]$ sector of the modulus-modulus response,
clearly visible in Fig.~\ref{fig:T0wp} d. Rather than a change in eigenenergy
or damping rate (note the relative flatness of $\omega_{\mathrm{pb}}$ and
$\gamma_{\mathrm{pb}}$, red curves in Figs.~\ref{fig:T0wp}a and \ref{fig:T0wp}%
e), this increase is caused by the transfer of the spectral weight of the
plasma branch (which is small but nonzero in the modulus-modulus channel) to
the former modulus mode, causing the emergence of a more intense mixed
modulus-plasma mode (note the increase of the residue modulus $\left\vert
Z_{II}^{(1)}\right\vert $ in Fig.~\ref{fig:T0wp} \emph{f}) . This enhancement
is promising for an experimental observation of modulus \textquotedblleft
Higgs\textquotedblright\ collective excitations in charged superfluids and superconductor.%

Fig.~\ref{fig:T0wp} $\left(  b,c\right)  $ also illustrates nicely the
splitting of the plasma resonance occurring in the range of values of
$\omega_{p}$ such that the bare plasma eigenfrequency lies within the interval
$[2\Delta,\omega_{2}^{(pp)}]$. Note that this effect is not due to an
interaction between the modulus and plasma branches, as it is also observed in
the BCS limit \cite{Plasma-PRL} where the two branches are decoupled.

\section{Vicinity of the phase transition \label{sec:tc}}

\begin{figure}[tbh]%
\centering
\includegraphics[
height=5.6342in,
width=4.4131in
]%
{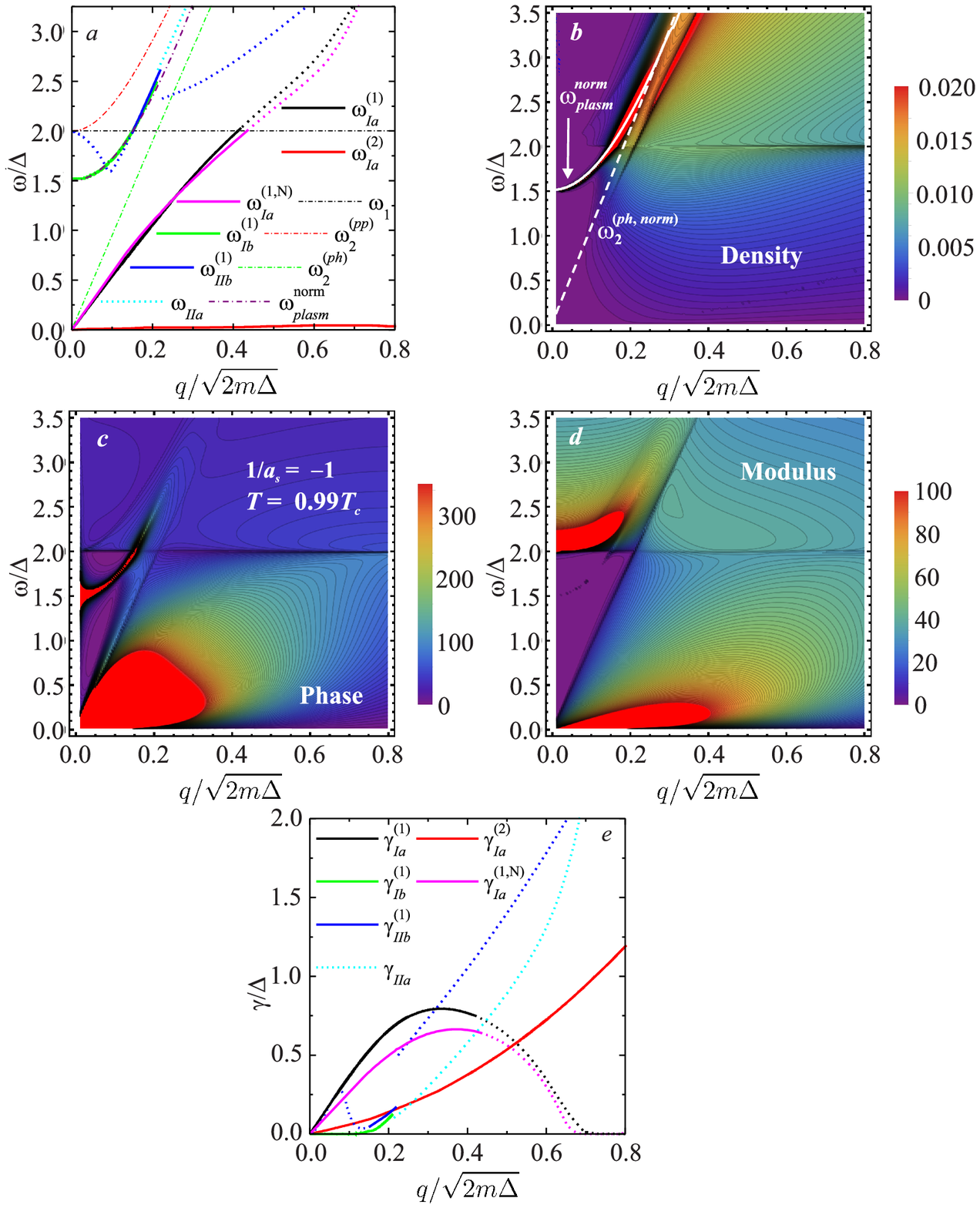}%
\caption{$(a)$ Eigenfrequencies provided by complex roots of $\det
\mathbb{K}_{\downarrow}$ (solid curves) at $1/k_{F}a_{s}=-1$, $\omega
_{p}=1.5\Delta$ and $T=0.99T_{c}$ as functions of $q$. $(b,c,d)$ Contour plots
of $\chi_{\rho\rho}$, $\chi_{\theta\theta}$ and $\chi_{\lambda\lambda}$ in the
same regime. $\left(  e\right)  $ Damping factors provided by imaginary parts
of the complex roots of of $\det\mathbb{K}_{\downarrow}$.White solid and
dashed curves in panel (\emph{b}) show, respectively, the plasma frequency and
the upper edge of the particle-hole continuum of the normal phase. }%
\label{fig:Tcq}%
\end{figure}

We now turn to the high-temperature regime $|T-T_{c}|\ll T_{c}$ which (as in
the neutral case \cite{AllModes-PRA}) differs much from the zero-temperature
case. Fig.~\ref{fig:Tcq} shows the dispersion and contour plots at
$T=0.99T_{c}$ and $1/k_{F}a=-1$ and $\omega_{p}=1.5\Delta=1.75\times
10^{-2}E_{F}$. We note that having $\omega_{p}$ comparable to $\Delta$ close
to the phase transition is not the typical experimental situation of
superconductors (where $\omega_{p}$ is rather fixed in units of $E_{F}$ so
large compared to $\Delta$ near $T_{c}$), this may be approximately the case
in highly-layered materials such as cuprates where at zero-temperature, the
plasma frequency in the transverse $c$-direction is much less than the gap.%

To understand the spectrum near $T_{c}$, one should first notice that density
fluctuations decouple from the pair-field fluctuations, in both modulus and
phase (in other words, $K_{1,3}$ and $K_{2,3}$ tends to 0 at $T_{c}$). This
reflects the situation in the normal phase where the pair susceptibility is
decoupled from the density-density response function. Thus, in $\chi_{\rho
\rho}$ (Fig.~\ref{fig:Tcq} \emph{b}), the plasma branch appears very close to
its normal limit (shown by the overlaid white solid curve in
Fig.~\ref{fig:Tcq} \emph{b}). Unlike at $T=0$, it is barely sensitive to the
structure of the pair-breaking continuum: no resonance splitting is visible
around $2\Delta$, $\omega_{2}^{(pp)}$ or $\omega_{3}$. Several roots of
$\det\mathbb{K}$ are associated to plasma modes in the analytic continuation
but they reconnect very well at the angular points (compare the green curve
$\omega_{Ib}^{(1)}$ and the blue curve $\omega_{IIb}^{(1)}$ in
Fig.~\ref{fig:Tcq} \emph{a}) indicating that they are physically equivalent.
Their remains however an important distortion of the plasma branch near
$\omega_{2}^{\mathrm{(ph)}}$ but this is not due to superconductivity: it is
nothing else than the distortion due to the upper-edge of the particle-hole
continuum of the normal phase. This edge is shown by the overlaid white dashed
curve in Fig.~\ref{fig:Tcq} \emph{b} and is approximated by $\omega
_{2}^{\mathrm{(ph)}}$ when $T$ is close to $T_{c}$. Overall, the re-emergence
of the normal plasma branch in the vicinity of the phase transition reflects
the loss of importance of the vanishingly small fraction of superconducting
electrons in carrying the plasma wave.

As it turns into its normal limit, the plasma resonance also looses its
spectral weight in the pair field channels (modulus and phase), which is
visible in Fig.~\ref{fig:Tcq}, \emph{c,d}. Within our approximation, Coulomb
interactions do not intrinsically enter into the pair-field propagator
$\mathbb{K}_{\mathrm{GPF}}$, such that above $T_{c}$ (when the $\rho$-$\Delta$
matrix elements $K_{1,3}$ and $K_{2,3}$ vanish) this propagator coincides with
the pair susceptibility of the neutral and normal Fermi gas with short range
interactions. We exclude here the exotic limiting case in which $\omega_{p}$
would tend to 0 with $|T-T_{c}|$ faster than $\Delta$ such that the plasma
frequency would remain comparable to that of the phononic branches. This
susceptibility (studied in \cite{AllModes-PRA}) displays a pairing mode with a
quadratic dispersion in the regime $1/\xi\ll q\ll k_{F}$ (with $\xi
\approx2m\Delta/k_{F}$ the pair coherence length), evolving into a double
phononic branch (corresponding to $\omega_{Ia}^{(1)}$ and $\omega_{Ia}^{(2)}$
in Fig.~\ref{fig:Tcq} \emph{a}) when $q\approx1/\xi$. A modulus mode near
$2\Delta$ also survives at wavelengths comparable to the pair coherence
length. Due to the decoupling from Coulomb interactions, this low-energy
collective physics emerges near $T_{c}$ as it would in the neutral case:
except for the residual plasma branch still appearing in the phase response,
the contour plots Fig.~\ref{fig:Tcq}, \emph{c} and \emph{d} nearly coincide
with their neutral equivalent. At nonzero temperature, the conjecture of
Anderson concerning the disappearance of phononic branches in charged systems
is thus limited to the density channel: in pair-field channels, a
\textquotedblleft collisionless second sound\textquotedblright%
\ \cite{AllModes-PRA} exists even in presence of Coulomb interactions. As can
be seen from Fig. \ref{fig:Tcq}, more than one gapless collective excitations
are predicted in the BCS-BEC crossover regime contrary to the far BCS limit
studied in preceding works. The dispersion of the higher-energy
Carlson-Goldman mode appears to be close to the phononic-like mode in a
neutral Fermi superfluid with the same scattering length \cite{AllModes-PRA}.
We recall that the other (low-velocity) gapless branch (red curve in
Fig.~\ref{fig:Tcq} \emph{a}) is computed here in the collisionless regime,
since we assumed that the fermionic quasiparticle have an infinite lifetime.
It is however reminiscent of the second gapless mode (gapless modes in charged
condensates are also called Carlson-Goldman modes \cite{Carlson}) derived
under hydrodynamic assumptions for the quasiparticle lifetime. This was
already noticed in \cite{Takada1} for the case of the BCS limit ($T_{c}\ll\mu
$), where this sound branch is visible only in a tiny temperature range below
$T_{c}$. We recall however that the critical behavior \cite{Popov1976,PB-PRA}
of the sound velocity is incorrectly predicted by Ref.~\cite{Takada1}.
Furthermore, we find no trace of the so-called \textquotedblleft upper
mode\textquotedblright\ introduced by the authors in the analytic
continuation. This mode is an artifact of solving the truncated dispersion
equation (see Eq. (2.25) in \cite{Takada1}) $\operatorname{Re}\left[
\det\mathbb{K}\left(  \omega+i0^{+}\right)  \right]  =0$ instead of the
equation (\ref{detKeq0a}). At stronger interactions, we note a broadening of
the visibility range.

\section{Limit of large plasma frequency}

In three-dimensional superconductors (as opposed to layered materials), the
concentration of carriers is typically rather high so that the plasma
frequency appears several orders larger than the pair-breaking continuum edge.
In the above subsections, we explored the regime when the plasma frequency is
low enough to be in resonance with pair-breaking and gapless modes. Here, on
the other hand, we focus on the regime of large $\omega_{p}$ compared to both
$\Delta$ and $T_{c}$. While this regime was previously considered only at
$q=0$ \cite{Anderson1958}, or by using large but finite numerical value of
$\omega_{p}/T_{c}$, we show here how to analytically take the limit
$\omega_{p}\to+\infty$ at fixed $q$, $T$ and $1/k_{F} a$. This regime
corresponds to most realistic superconductors and therefore is relevant for
contemporary experiments.


In the limit $\omega_{p}\rightarrow+\infty$ and for complex frequencies low
compared to the plasma frequency ($|z|\ll\omega_{p}$), one should redefine the
inverse fluctuation propagator as
\begin{equation}
\mathbb{\tilde{K}}=\left(
\begin{array}
[c]{ccc}%
K_{1,1} & K_{1,2} & \tilde{K}_{1,3}\\
-K_{1,2} & K_{2,2} & \tilde{K}_{2,3}\\
\tilde{K}_{1,3} & -\tilde{K}_{2,3} & \tilde{K}_{3,3}%
\end{array}
\right)  \label{Ktild}%
\end{equation}
where the rescaled matrix elements
\begin{equation}
\tilde{K}_{1,3}=\frac{1}{\sqrt{\alpha_{0}}}K_{1,3},\quad\tilde{K}_{2,3}%
=\frac{1}{\sqrt{\alpha_{0}}}K_{2,3},\quad\tilde{K}_{3,3}=\frac{1}{\alpha_{0}%
}\left(  K_{3,3}-\frac{q^{2}}{4\pi}\right)  , \label{mels}%
\end{equation}
have a finite and nonzero limit when $\omega_{p}\rightarrow+\infty$ (we recall
that $\alpha_{0}=3\pi\omega_{p}^{2}/8$).

The dispersion equation then transforms to
\begin{equation}
\lim_{\alpha_{0}\rightarrow\infty}\left(  \frac{1}{\alpha_{0}}\det
\mathbb{K}\right)  =\det\mathbb{\tilde{K}}, \label{limK}%
\end{equation}
Also the spectral weight functions (\ref{XLL}), (\ref{XYY}), (\ref{RF1}) for
$\left(  \omega,\Delta\right)  \ll\omega_{p}$ are determined by the
expressions,%
\begin{align}
\lim_{\alpha_{0}\rightarrow\infty}\chi_{\lambda\lambda}\left(  \mathbf{q}%
,\omega\right)   &  =\frac{1}{\pi}\operatorname{Im}\frac{K_{2,2}\left(
\mathbf{q},\omega+i0^{+}\right)  \tilde{K}_{3,3}\left(  \mathbf{q}%
,\omega+i0^{+}\right)  +\tilde{K}_{2,3}^{2}\left(  \mathbf{q},\omega
+i0^{+}\right)  }{\det\mathbb{\tilde{K}}\left(  \mathbf{q},\omega
+i0^{+}\right)  },\label{XLLb}\\
\lim_{\alpha_{0}\rightarrow\infty}\chi_{\theta\theta}\left(  \mathbf{q}%
,\omega\right)   &  =\frac{1}{\pi}\operatorname{Im}\frac{K_{1,1}\left(
\mathbf{q},\omega+i0^{+}\right)  \tilde{K}_{3,3}\left(  \mathbf{q}%
,\omega+i0^{+}\right)  -\tilde{K}_{1,3}^{2}\left(  \mathbf{q},\omega
+i0^{+}\right)  }{\det\mathbb{\tilde{K}}\left(  \mathbf{q},\omega
+i0^{+}\right)  },\label{XYYb}\\
\lim_{\alpha_{0}\rightarrow\infty}\chi_{\rho\rho}\left(  \mathbf{q}%
,\omega\right)   &  =-\frac{q^{4}}{16\pi^{3}}\operatorname{Im}\left(
\frac{\det\mathbb{K}_{GPF}\left(  \mathbf{q},\omega+i0^{+}\right)  }%
{\det\mathbb{\tilde{K}}\left(  \mathbf{q},\omega+i0^{+}\right)  }\right)  ,
\label{RF1b}%
\end{align}
which are independent of $\alpha_{0}$ far from the plasma resonance, as expected.%

\begin{figure}[ptbh]%
\centering
\includegraphics[
height=4.4287in,
width=3.2344in
]%
{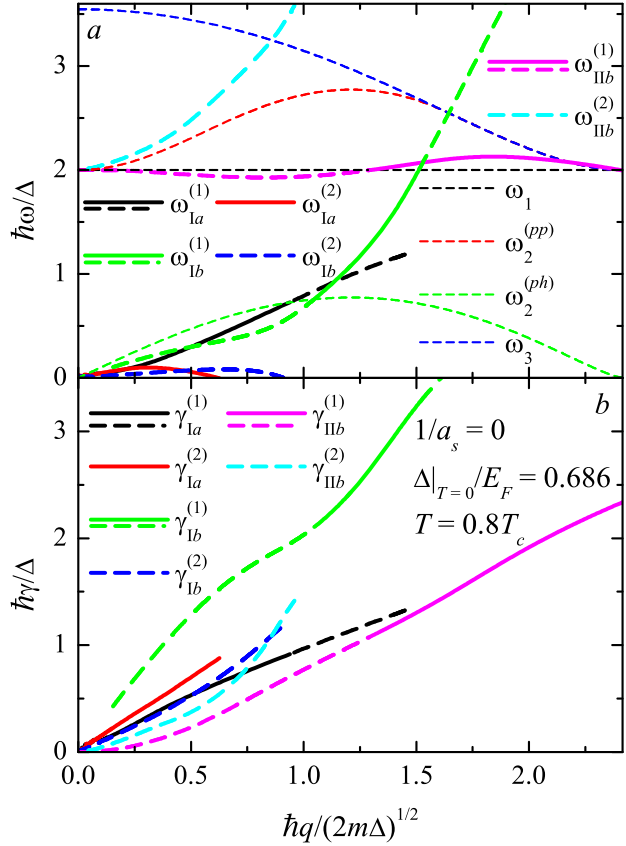}%
\caption{Dispersion (\emph{a}) and damping (\emph{b}) of low-lying collective
excitations for a charged superfluid at unitarity, $1/a_{s}=0$ ($\left.
\Delta\right\vert _{T=0}\approx0.686E_{F}$) at the temperature $T=0.8T_{c}$.
The notations are the same as in Fig. \ref{fig:T0q}.}%
\label{fig:CGPoles}%
\end{figure}

Since the double limit $\omega_{p}\rightarrow+\infty$ and $1/k_{F}%
a\rightarrow-\infty$ (\textit{i.e.} the BCS limit) was already discussed in
the literature \cite{Takada1}, we show in Fig. \ref{fig:CGPoles} the
collective excitations in the strong-coupling regime (more precisely at
unitarity $1/k_{F}a=0$). At the considered temperature, the excitation
spectrum resembles the neutral spectrum, with both high- and low-velocity
gapless branches (respectively black and red lines), a pair-breaking branch
around $2\Delta$. The main difference with the neutral case is $\omega_{IIb}^{(2)}$, 
which, as in the case of $\omega_{p}$ finite, is responsible
for a peak near $2\Delta$ in the phase and density responses. This peak is
absent in the neutral case and therefore characteristic of the charged system.
We can see also that the gapless Carlson-Goldman branches can survive at
strong coupling in a wider temperature range below $T_{c}$ with respect to the
BCS case. Moreover, the lower-energy gapless mode is better resolved at
stronger coupling.

\begin{figure}[ptbh]%
\centering
\includegraphics[
height=4.3258in,
width=2.9724in
]%
{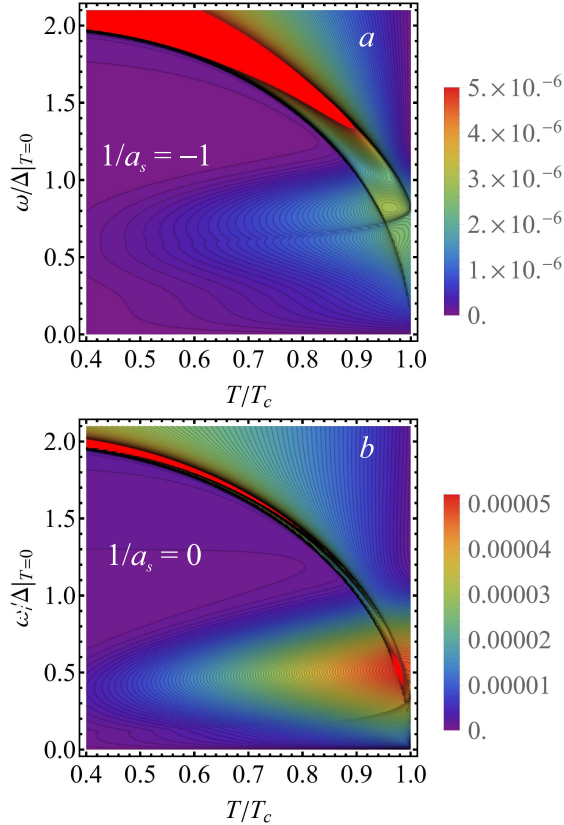}%
\caption{Contour plots of the spectral functions for the density-density
response in the low-frequency region with respect to the plasma frequency when
varying temperature at a fixed momentum $\hbar q=0.2\sqrt{2m\left.
\Delta\right\vert _{T=0}}$ in the BCS regime with $1/k_{F}a_{s}=-1$ (\emph{a})
and at unitarity (\emph{b}).}%
\label{fig:CarlsonContPlots}%
\end{figure}

Fig. \ref{fig:CarlsonContPlots} shows the temperature evolution of
density-density response at a fixed momentum $\hbar q=0.2\sqrt{2m\left.
\Delta\right\vert _{T=0}}$ for the BCS regime with $1/k_{F}a_{s}=-1$ and at
unitarity. At low temperature, the peak linked to $\omega_{IIb}^{(2)}$ is well
visible in the interval $[2\Delta,\omega_{2}^{(pp)}]$. When approaching
$T_{c}$, this peak disappears and instead a broad feature corresponding to the
normal density response develops at energies comparable to $\epsilon_{F}$.%

\section{Conclusions \label{Conclusions}}

In this paper, we investigated theoretically the dispersion and damping of
collective excitations in charged and condensed Fermi gases at finite
temperatures and in the BCS-BEC crossover. The treatment is performed within
the Gaussian pair and density fluctuation (GPDF) method. The spectra of
collective excitations are considered using two complementary methods: (i)
exploration of the density-density, phase-phase and modulus-modulus response
functions, (ii) determination of eigenfrequencies and damping factors of
collective excitations from complex poles of the GPDF propagator analytically
continued through its branch cuts. Comparison of results obtained by these two
methods gives us a reliable and detailed picture of collective excitations.

At zero temperature, two collective excitation branches dominate the spectrum:
the plasma mode and the pair-breaking mode. The gapped plasma branch
continuously evolves to the gapless phononic branch of the neutral condensate
as Coulomb interaction are turned off (that is, as the plasma frequency is
sent to zero). When the plasma mode crosses the pair-breaking continuum edge,
it exhibits a resonant anticrossing with the pair-breaking mode.
Correspondingly, the magnitude of the modulus-modulus spectral function,
greatly increases on resonance, which can facilitate the experimental
detection of pair-breaking modes.

At nonzero temperatures, a phononic branch may still exists in the charged
system, coexisting with the gapped plasma and pair-breaking branch. This mode
describes the motion of a minority of superconducting electrons in a majority
of normal carrier. In preceding works devoted to BCS superconductors, it was
shown to only existed in a close vicinity of $T_{c}$. At stronger couplings,
in the BCS-BEC crossover, the conditions are substantially more favorable for
survival of the {\normalsize gapless} mode.

The spectra of collective excitations considered in the present work can be a
subject of experimental verification in spectral measurements of the density
response of charged superfluids. The pair field response can also be
experimentally detected, e.~g. by tunneling experiments, similar to the
Carlson-Goldman experiment \cite{Carlson}. Our method can easily be transposed
to 2D or quasi-2D systems of condensed charged fermions, which makes it
promising for the treatment of collective excitations in layered and
high-$T_{c}$ superconductors.

\begin{acknowledgments}
We acknowledge funding by the Research Foundation -- Flanders, projects
GOH1122N, G061820N, G060820N, and by the University Research Fund (BOF) of the
University of Antwerp.
\end{acknowledgments}

\end{document}